\documentclass[article]{aa}
\usepackage{txfonts}
\usepackage{graphicx}
\usepackage{multirow}

\usepackage{natbib}

\begin{document}

\title{On the redshift of the bright BL Lac object PKS 0048-097} 

\author{M. Landoni \inst{1,2}  
     \and R. Falomo\inst{3}
     \and A. Treves \inst{1,4}
     \and B. Sbarufatti\inst{2} 
     \and R. Decarli \inst{5}
     \and F. Tavecchio \inst{2}
     \and J. Kotilainen \inst{6}
     }

\institute{Universit\`{a} degli Studi dell'Insubria.
Via Valleggio 11, I-22100 Como, Italy. 
  \and INAF - Osservatorio Astronomico di Brera, Via Bianchi 46, I–23807 Merate (LC), Italy
  \and INAF - Osservatorio Astronomico di Padova, Vicolo dell' Osservatorio 5, I-35122 Padova, Italy
  \and INAF - Istituto Nazionale di Astrofisica and INFN - Istituto Nazionale di Fisica Nucleare
  \and Max-Planck-Institut fur Astronomie, Konigstuhl 17, 69117 Heidelberg, Germany
  \and Finnish Centre for Astronomy with ESO (FINCA) - University of Turku, V\"ais\"al\"antie
 20, FI-21500 Piikki\"o, Finland}

\date{Received 25 February 2012 / Accepted 26 April 2012}

\abstract {} { The determination of elusive redshifts of bright BL Lac objects} {We use the capabilities of newly available spectrograph X-Shooter at European Southern Observatory (ESO) Very Large Telescope, that combines high resolution and a large wavelength range, to obtain UVB to near-IR spectra of BL Lacs.}
{Our observations of PKS 0048-097 detect three emission lines that permit to derive a redshift $z = 0.635$. Moreover, a Mg II absorption system at $z = 0.154$ that is associated with a foreground spiral galaxy at 50 Kpc projected distance is found.} {The obtained redshift allows us to comment about the optical beaming factor and the absorption of the high energy spectrum by the Extragalactic Background Light.}

{
\maketitle 

\section{Introduction}
PKS 0048-097 ($m_{V}$ $\sim$ 15.5) is a bright, strongly variable \citep{pica83} and well studied BL Lac object \citep{wills92, fal93, fal94,stickel93}. 
Despite many spectroscopical studies \citep{stickel93, sba06} it remains one of the few bright sources with unknown redshift. Imaging studies of the source were unable to detect its host galaxy \citep{fal96}. Moreover, the upper limits to the brightness of the host galaxy derived from optical and NIR imaging suggest lower limits to the redshift of $z > 0.3$, even $z > 0.5$ \citep{fal96, kot98}. 
  
Stimulated by the availability of new spectroscopical instrumentation (ESO X-Shooter, \citet{vernet11}) we collected optical to near-IR spectra of the object aimed at detecting very faint features over a large spectral range. 

The paper is organized as follow. Observations and data reduction are described in Section 2. Main spectroscopical results are reported in Section 3 where we also comment on Mg II intervening system that required a reconsideration of the close enviroment of the BL Lac. 
Discussion and conclusions are given in Section 4.
Throughout the paper, we consider the following cosmological parameters: $H_{0} = 70$ km s$^{-1}$    Mpc$^{-1}$, $\Omega_{m} = 0.27$ and $\Omega_{\Lambda} = 0.73$.

\section{Observations and data reduction}
 \begin{figure*}
\centering
   \includegraphics[width=18.5cm]{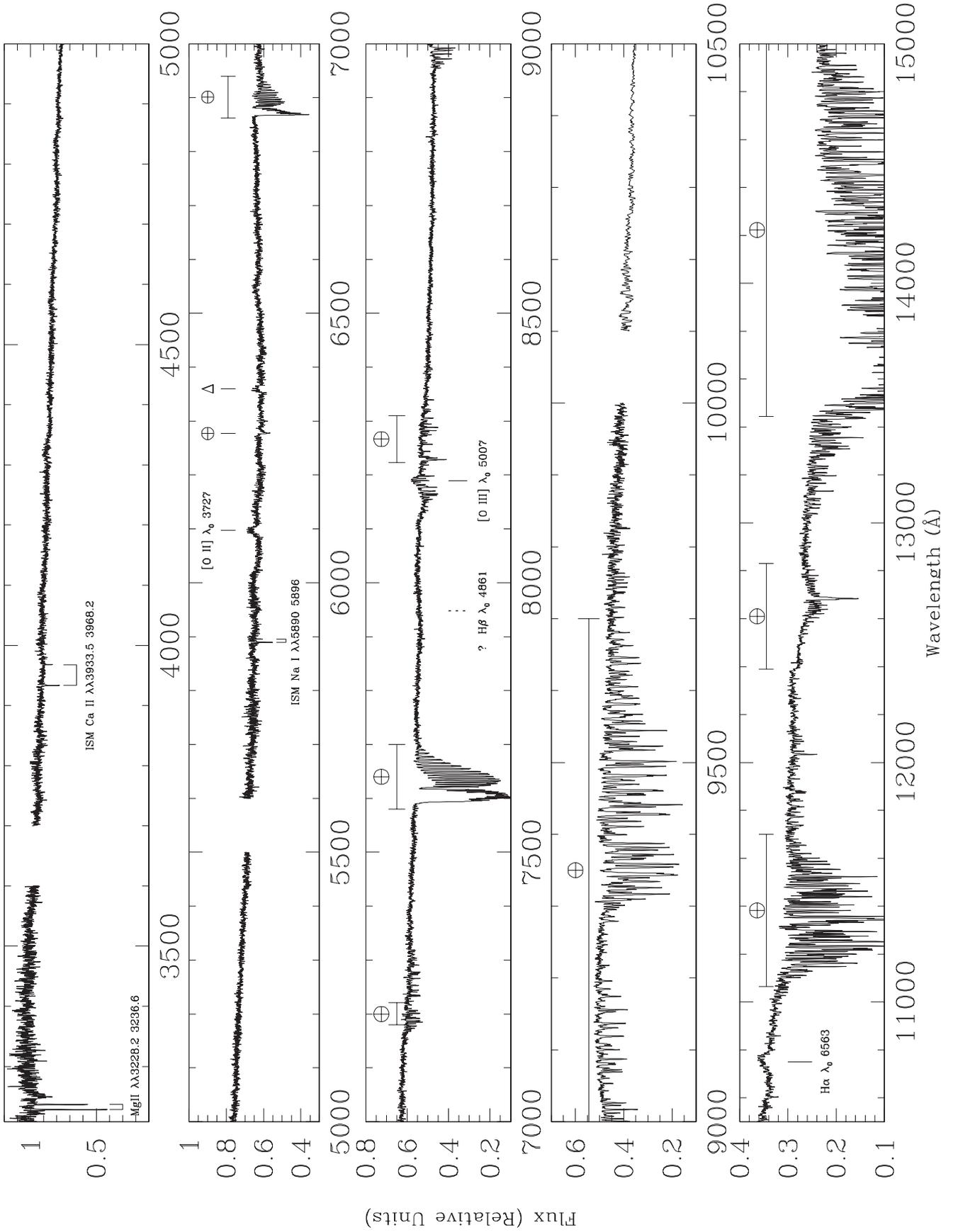}
     \caption{PKS0048-097 overall mean spectrum. Atmospheric absorption are labelled by $\oplus$ while calibration artefacts (detectable on the 2D image of the spectrum) are marked with $\Delta$. Emission lines above $\sim 3\sigma$ signal-to-noise ratio of the spectrum are marked with a single vertical line. The expected position of H$\beta$ at $z = 0.635$ is marked with a dashed line. }
     \label{fig:meanspec}
\end{figure*}

We obtained UVB (3100-5500 $\AA$), Optical (5600-10.000 $\AA$) and near-IR (10.100-15.000 $\AA$) spectra of the target using the ESO Very Large Telescope (UT2 - \textit{Kueyen}) equipped with X-Shooter in Service Mode.
Two independent spectra were secured using the Nodding configuration for the instrument (see Table ~\ref{table:jobs} for detailed information).

\begin{table*}
\caption{Journal of Observations of PKS0048-097}              
\label{table:jobs}      
\centering     
\begin{tabular}{c c c c c c c c}
\hline\hline 
  Date of Observation \tablefootmark{a} & Seeing \tablefootmark{b} & Channel \tablefootmark{c} & Slit Width \tablefootmark{d}& R \tablefootmark{e}& $t_{exp}$ \tablefootmark{f} & $N$ \tablefootmark{g} & $S/N$ \tablefootmark{h} \\
  \hline

 &  & UVB & $1.6^{\prime\prime} \times 11^{\prime\prime}$ & 3300 & 2720 & 4 & 34\\

05 Jul 2010 & 1.22 & VIS & $1.5^{\prime\prime} \times 11^{\prime\prime}$ & 5400  & 2460 & 6 &32\\
  & & NIR &$1.5^{\prime\prime} \times 11^{\prime\prime}$ & 3500  & 1440 & 6 & 28\\

  \hline

 &  & UVB & $1.6^{\prime\prime} \times 11^{\prime\prime}$ & 3300  & 2720 & 4 & 24 \\
18 Aug 2010 &0.84 & VIS & $1.5^{\prime\prime} \times 11^{\prime\prime}$ & 5400  & 2460 & 6 & 24 \\
  & & NIR & $1.5^{\prime\prime} \times 11^{\prime\prime}$ & 3500  & 1440 & 6  & 24\\

  \hline
\end{tabular}
\tablefoot{
The top panel shows the Journal of Observations for the bright BL Lac object PKS0048-097.
\tablefoottext{a}{Date of Observation}
\tablefoottext{b}{Seeing during the observation.}
\tablefoottext{c}{X-Shooter Spectrograph arm.}
\tablefoottext{d}{Used slit for the observation (in arcsec)}
\tablefoottext{e}{Resolution ($\lambda / \delta\lambda$)}
\tablefoottext{f}{Time of exposure (in seconds).}
\tablefoottext{g}{Number of spectra obtained in NOD mode.
\tablefoottext{h}{Average $S/N$ ratio for the channel}
}
}

\end{table*}

Data reduction was carried out using the X-Shooter Data Reduction Pipeline (version 1.3.7, see \citet{goldoni11}) in Polynomial Mode adopting the configuration recommended by ESO.
In particular, for each observation, we calculated the Master Bias Frames and Master Flat Frames for each channel selecting carefully the proper slit values (see Table ~\ref{table:jobs}). We also obtained the bidimensional mapping required by the pipeline stack to resample the Echelle Orders. Finally, for each channel, we computed the sensitivity functions required by the pipeline to calibrate in flux the spectra. We normalized in relative units the obtained spectra in order to proper juxtapose the three arms. 
The reduction of the data and the next extraction and calibration of the spectrum was performed for the two individual exposures (see above). 
Since no significant spectral differences were apparent between the two spectra we averaged them in order to improve the signal-to-noise ratio.

In Figure~\ref{fig:meanspec} we show the mean spectrum of PKS0048-097.
The continuum is clearly dominated by the contribution of non-thermal emission which can be described by a  power-law of the form $F(\lambda) = {\lambda}^{-\alpha}$ ($\alpha \sim 0.9$). In addition to many telluric absorption lines we clearly detect intervening Galactic absorption of Ca II (EW: 0.16 $\AA$ - 0.05 $\AA$) and Na I (EW: 0.10 $\AA$ - 0.07 $\AA$).
The flux calibration has been assessed throught the relative photometric calibration of the acquisition image. The R band magnitude of PKS 0048-097 is found to be $17.3 \pm 0.3$ which corresponds to a low state of the source. In fact, in the literature the R magnitude of the source is reported between 14.74 and 16.22 (\citet{fan2000} see also REM archives, Angela Sandrinelli private communication). The combination of the high quality instrumentation, in terms of throughput and spectral resolution, and the low state of the source clearly favored the detection of the faint emission lines.

\section{Results}

\begin{figure}
  \resizebox{\hsize}{!}{\includegraphics{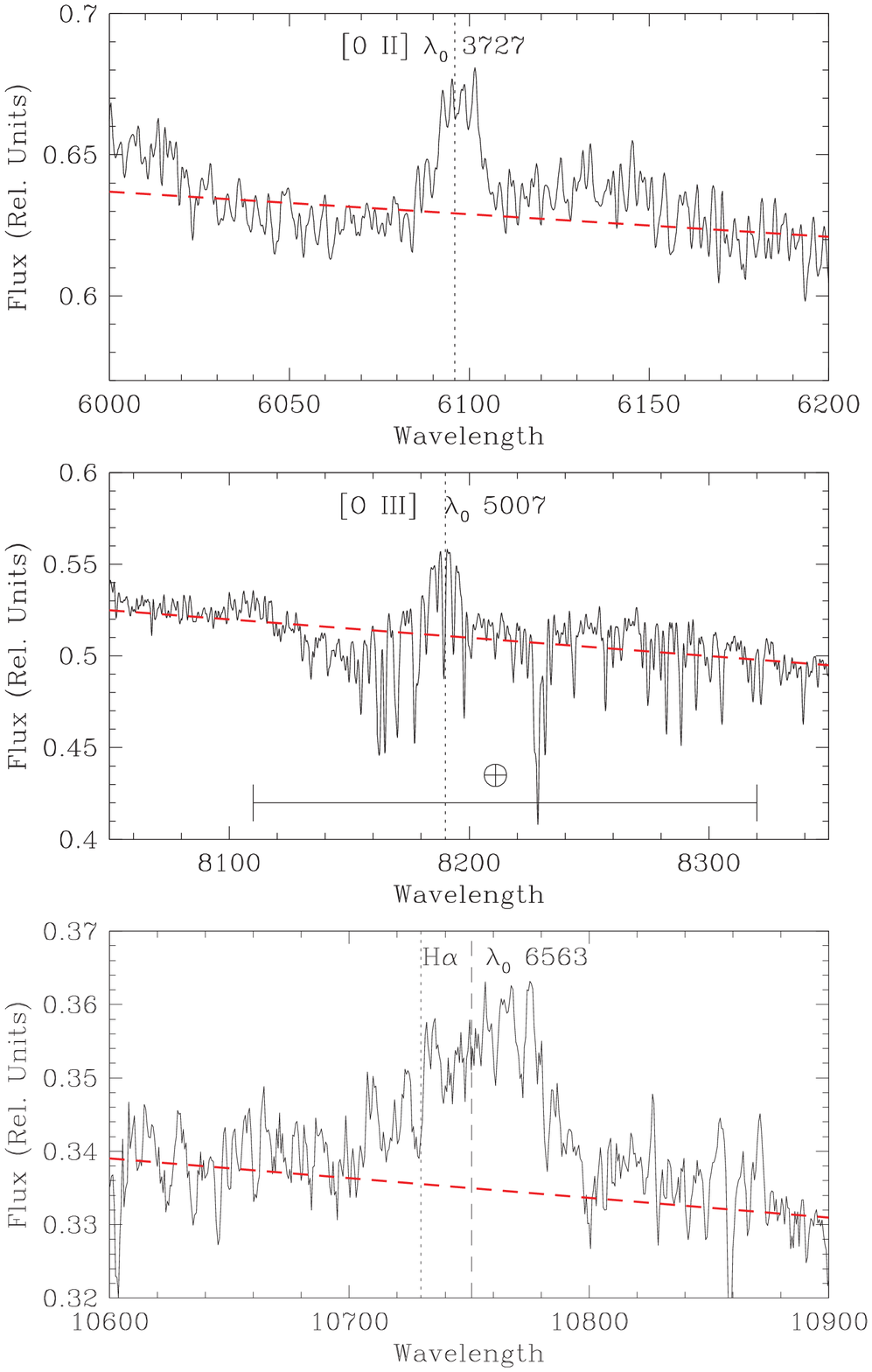}}
  \caption{PKS 0048-097 [OII] emission line at 6096 $\AA$, [O III] at 8189 $\AA$ polluted by many resolved atmospheric absorption and H $\alpha$ emission line at 10751 $\AA$. The vertical dotted lines represent the expected position of the three emission lines assuming the redshift z = 0.635, while the dashed one indicate the position of barycenter. Red dotted line is the fit of the continuum.}
       \label{fig:emissions}
\end{figure}

\begin{figure}
\centering
  \resizebox{\hsize}{!}{\includegraphics{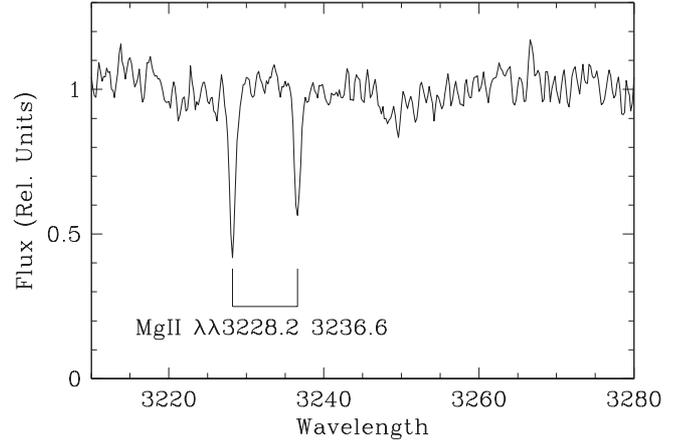}}
  \caption{PKS 0048-097 Mg II absorption features at $z = 0.154$ associated with a spiral galaxy at 50 Kpc projected distance from the BL Lac.}
      \label{fig:mgiisystem}
\end{figure}

 \begin{figure*}

   \includegraphics[width=18cm]{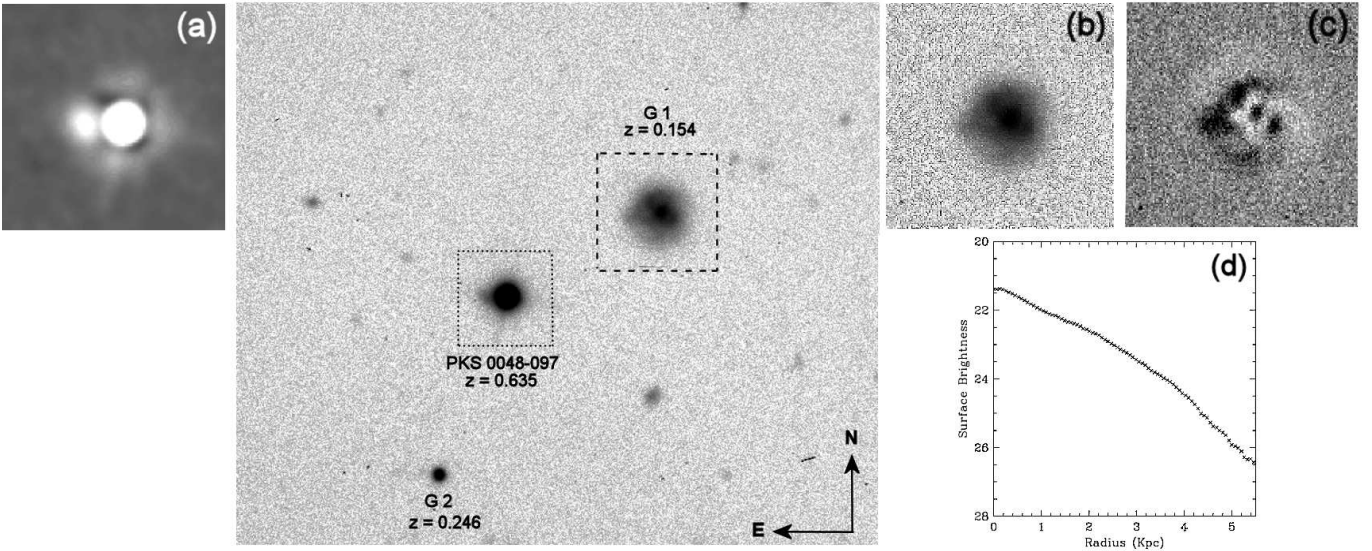}
     \caption{R Band image (FoV of $1.1^{\prime} \times 1.1^{\prime}$) of the BL Lac object PKS 0048-097 obtained with NTT and SUSI \citep{european1994emmi} by \citet{fal96} that shows its close enviroment. The spiral galaxy $G1$ (panel b) is the Mg II intervening system detected in the very blue part of the spectrum while the Galaxy G2 is a foreground object respect to the absorption. The panel (a) in the figure shows a deconvolution of the image. The small object nearby PKS 0048-097 is most probably a background galaxy. Panel (c) shows the image of galaxy G1 after the model subtraction while the panel (d) reports the average radial brightness profile of galaxy G1.}
     \label{fig:fov}
\end{figure*}

\begin{table*}
\caption{Measurements of spectral lines observed (referred to the mean spectrum of each arm). The observed $\lambda$ column is referred to the barycenter of the line.}              
\label{table:lines}      
\centering                                      
\begin{tabular}{c l l l c c}          
\hline\hline                        
\phantom{|}Line ID  & \phantom{aaaa}Observed $\lambda$ &\phantom{aaaaaa} z & Observed  FWHM   & Observed EW  & Notes  \\     

& \phantom{aaaaaaa.}$[\AA]$  & &\phantom{aaaa}$[$km s$^{-1}]$ & $[\AA]$ &    \\ 
\hline                                   
    Mg II\phantom{-}  2796.35   & \phantom{.aa}3228.20 $\pm$ 0.20 & 0.1544 $\pm$ 0.0001  & & 0.59 $\pm$ 0.16 & abs. \\      
    Mg II\phantom{-}   2803.53 &\phantom{aa} 3236.60 $\pm$ 0.20 & 0.1544 $\pm$ 0.0001  & & 0.45 $\pm$ 0.12 &abs. \\ 

    $[$O II$]$\phantom{-}  3727.40  &\phantom{aa} 6096.20 $\pm$ 1\phantom{.00} & 0.635 \phantom{..}$\pm$ 0.0004  &\phantom {aaa.}680\phantom{} $\pm$ 100 & 0.79 $\pm$ 0.39 &em. \\
    $[$O III$]$  5006.84  & \phantom{aa} 8189.20 $\pm$ 1\phantom{.00} & 0.635\phantom{..} $\pm$ 0.0004  &\phantom {aaa.}400\phantom{} $\pm$ 100 & 0.92 $\pm$ 0.44 &em. (polluted by telluric absorptions) \\
    H$\alpha$\phantom{-}   \phantom{-}   \phantom{-}     6562.80  &\phantom{ .}10751.10\phantom{.} $\pm$ 5\phantom{.00} & 0.638\phantom{...}$\pm$ 0.001  & \phantom{aa.}1900\phantom{.}$\pm$ 130 & 3.00 $\pm$ 1.20 &em. \\

\hline                                             
\end{tabular}

\end{table*}
We searched for intrinsic emission or absorption features above $\sim 3\sigma$ signal-to-noise ratio of the spectrum (see Table ~\ref{table:jobs}}) with the additional condition that the features are clearly detectable in both individual spectra and avoiding obvious defects in the 2D image. This procedure brings three emission lines (see Figure ~\ref{fig:meanspec}, ~\ref{fig:emissions} and Table ~\ref{table:lines}) at $\lambda = 6096 \AA$ , $\lambda = 8189 \AA$ and $\lambda = 10751 \AA$. The line at $\lambda = 6096 \AA$ was tentatively detected by \citet{rector01} (EW: 0.3 $\AA$). We note that no previous near-IR observation of comparable spectral resolution is present in the literature.
We identify the first two lines as [O II] ($\lambda_{0} = 3727 \AA$) and [O III] ($\lambda_{0} = 5007 \AA$) from which a consistent redshift of $z = 0.635$ is derived. The third emission line ($\lambda = 10751 \AA$) is identified as H$\alpha$ that corresponds to $z = 0.638$. The small velocity difference ($\sim 600$ km s$^{-1}$) between the redshift derived from the narrow emission lines and the broad ones is not uncommon in quasars (see e.g. \citet{boroson05}). We therefore assume that the redshift of PKS 0048-097 is consistently $z = 0.635$. 
\\

We detect two narrow absorption features at $\lambda\lambda$ 3228.20 $\AA$ 3236.60 $\AA$ (see Figure ~\ref{fig:mgiisystem}) that we identify as an intervening Mg II absorption system at redshift $z = 0.154$. In order to search for objects associated with this absorption we investigated the enviroment around the target using high resolution deep image of the field, in the R band, obtained with ESO NTT and SUSI \citep{fal96}.  The close companion galaxy ($m_{R} = 22.5$) at $2.5^{\prime\prime}$ East of PKS 0048-097 (see panel (a) of Figure ~\ref{fig:fov}) already noted by \citet{fal90} is too faint to be associated with the absorption system at $z = 0.154$ and it is likely a background galaxy. On the other hand, the spiral galaxy G1 ($m_{r} = 19.2$) at $\sim 23^{\prime\prime}$ North-West from the BL Lac corresponding to 50 Kpc projected distance at redshift $z = 0.154$ \citep{stickel93} is the obvious galaxy associated to the observed Mg II absorption. This galaxy ( $M_{r} = -20.57$) has a disturbed morphology (see panel (b), (c) and (d) of Figure ~\ref{fig:fov}) after model subtraction and a scale length of $\sim 5$ Kpc. This is a rare case-study of absorption system at low redshift (see e.g. \citet{kac11}) in which the counterpart can be well  investigated through imaging analysis.
The column density correlated to the Mg II features, calculated adopting the Voigt profile fitting method, is   $\log N_{Mg II}^{a} = 13.20$.

\section{Beaming and Spectral Energy Distribution of PKS 0048-097}

\begin{figure}
\centering
  \resizebox{\hsize}{!}{\includegraphics{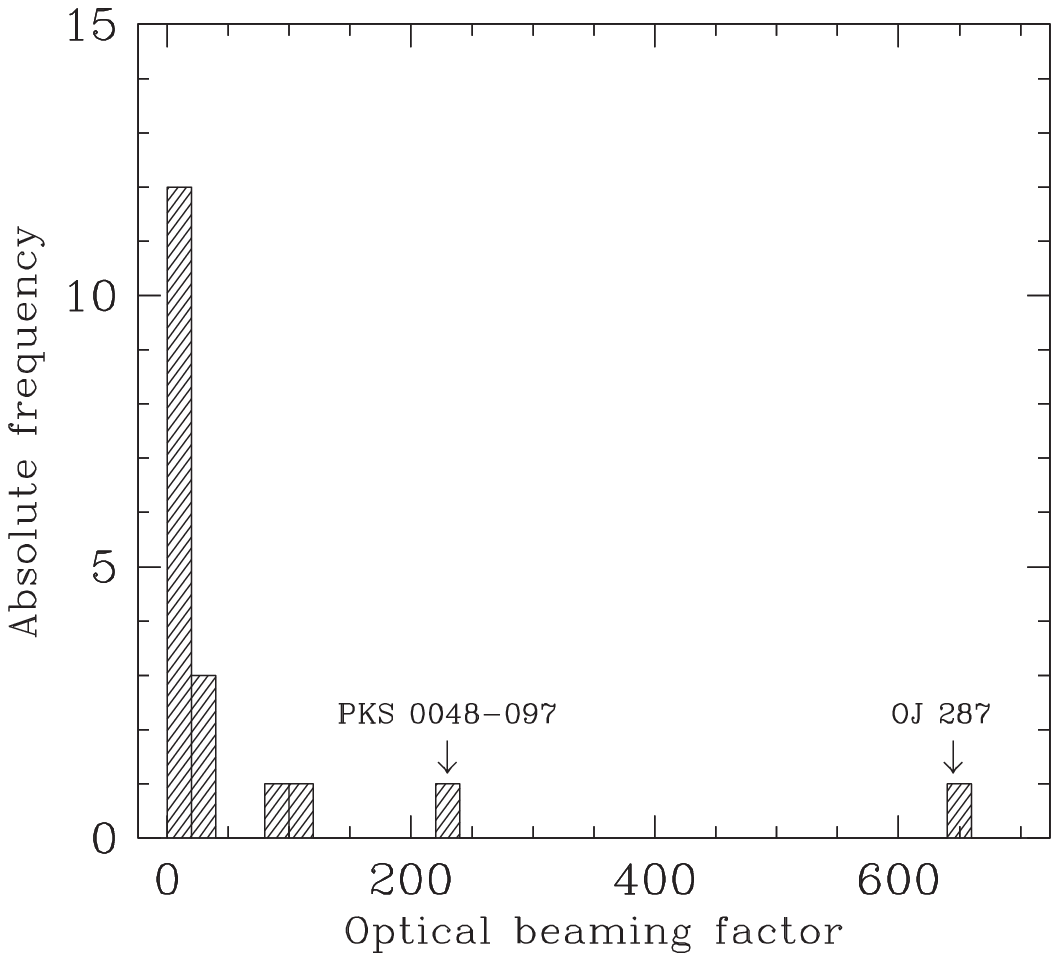}}
  \caption{ Distribution of Optical Beaming Factors for the BL Lac objects and PKS 0048-097. The extreme values for OJ 287 and PKS 0048-097 are marked by a down arrow.}
      \label{fig:beamfact}
\end{figure}

The luminosity of H$\alpha$ ($\log H\alpha \sim 41.93$ erg s$^{-1}$) of PKS 0048-097 is comparable with that of other BL Lac objects \citep{decarli11}. 
In order to derive the Optical Beaming Factor, defined as the ratio between $L_{c}$ and $L_{th}$, we calculated the expected thermal luminosity of the continuum at $5100 \AA$ ($\log L_{th} \sim 43.16$ erg s$^{-1}$, see \citet{decarli11}) and the continuum luminosity at the same wavelength ($\log L_{c} \sim 45.53$ erg s$^{-1}$). In the case of PKS 0048-097 this ratio is $\sim 234$.  
The extreme value of this beaming (second only to OJ 287, see Figure ~\ref{fig:beamfact}) is consistent with no detection of the host galaxy \citep{fal96}.

\begin{figure}
  \resizebox{\hsize}{!}{\includegraphics{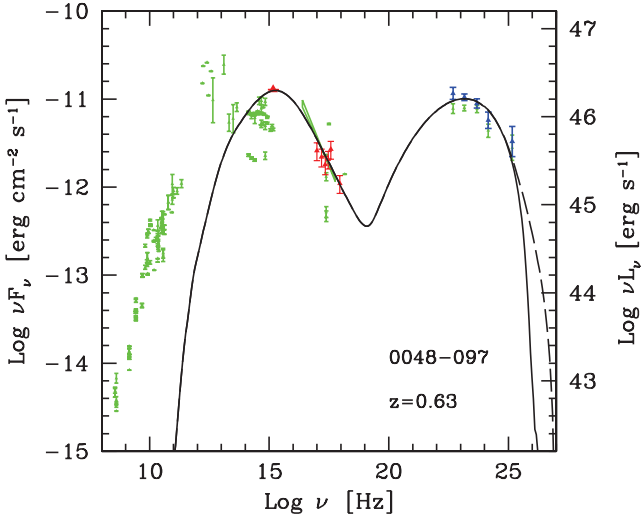}}
  \caption{PKS 0048-097 Spectral Energy Distribution. The dashed line is the intrinsic SED model while the solid line is the absorbed by EBL interaction. Green points are from ASDC archive. Red points represent Swift UVOT and XRT data. FERMI spectra FGL1 and FGL2 are indicated respectively with blue and green points. The parameters of the model are $\gamma _{\rm min}=500$, $\gamma _{\rm b}=8.5\times 10^3$ and $\gamma _{\rm max}=5\times 10^5$, $n_1=2$, $n_2=4.1$, $B=0.46$ G,  $K=3.7\times 10^4$, $R=8.2\times 10^{15}$ cm and $\delta=25$. For the references to the observations and details on the model see \citet{tav10}. }
       \label{fig:0048sed}
\end{figure}

PKS 0048-097 is a $\gamma$-ray emitter observed with FERMI in the band 100 MeV - 10 GeV and it is well fitted by a single power law ($\alpha_{0} = 2.38 \pm 0.24$). Assuming the extrapolation of this power law to the TeV band we calculated the expected emission corrected for the EBL absorption \citep{domin11} adopting the derived redshift. The EBL interaction reduces the intrinsic flux above 50 GeV by a factor 1.7 and above 300 GeV by a factor 60.
In Figure ~\ref{fig:0048sed} we reported the overall SED of PKS0048-097 collected for multifrequency data by \citet{tav10}. We fitted the data using a standard leptonic one-zone model considering synchrotron and synchrotron self-Compton (SSC) radiation mechanism (see \citet{maraschi2003}). We report two curves: the dashed line shows the intrinsic emission, while the solid one takes into account absorption by EBL photons, assuming the estimate of \citet{domin11}. We recall that, despite different approaches and methods, models for the EBL
substantially agree in the spectral region interesting for absorption of GeV-TeV
photons (see discussion and comparison in \citet{domin11}).
The one-zone models satisfactorily reproduce the SED above the IR band,
below which the emission is self-absorbed in the region we consider. The emission at
lower frequencies is therefore produced in more distant, transparent, region of the
jets not relevant for the modeling of the high-frequency SED. These consideration may be relevant in discussing the observability of the source with Cherenkov telescopes.

\begin{acknowledgements} 
We are grateful to P. Goldoni and E. P. Farina for comments.
\\
Part of this work is based on archival data services provided by the  
ASI Science Data Center ASDC.
\end{acknowledgements}

\bibpunct{(}{)}{;}{a}{}{,} 
\bibliographystyle{aa} 

\bibliography{biblio}{}

\begin{thebibliography}{22}
\expandafter\ifx\csname natexlab\endcsname\relax\def\natexlab#1{#1}\fi

\bibitem[{{Bonning} {et~al.}(2007){Bonning}, {Shields}, \&
  {Salviander}}]{bonning07}
{Bonning}, E.~W., {Shields}, G.~A., \& {Salviander}, S. 2007, \apjl, 666, L13

\bibitem[{{Boroson}(2005)}]{boroson05}
{Boroson}, T. 2005, \aj, 130, 381

\bibitem[{{Decarli} {et~al.}(2011){Decarli}, {Dotti}, \& {Treves}}]{decarli11}
{Decarli}, R., {Dotti}, M., \& {Treves}, A. 2011, \mnras, 413, 39

\bibitem[{{Dom{\'{\i}}nguez} {et~al.}(2011){Dom{\'{\i}}nguez}, {Primack},
  {Rosario}, {Prada}, {Gilmore}, {Faber}, {Koo}, {Somerville},
  {P{\'e}rez-Torres}, {P{\'e}rez-Gonz{\'a}lez}, {Huang}, {Davis},
  {Guhathakurta}, {Barmby}, {Conselice}, {Lozano}, {Newman}, \&
  {Cooper}}]{domin11}
{Dom{\'{\i}}nguez}, A., {Primack}, J.~R., {Rosario}, D.~J., {et~al.} 2011,
  \mnras, 410, 2556

\bibitem[{{Falomo}(1996)}]{fal96}
{Falomo}, R. 1996, \mnras, 283, 241

\bibitem[{{Falomo} {et~al.}(1993){Falomo}, {Bersanelli}, {Bouchet}, \&
  {Tanzi}}]{fal93}
{Falomo}, R., {Bersanelli}, M., {Bouchet}, P., \& {Tanzi}, E.~G. 1993, \aj,
  106, 11

\bibitem[{{Falomo} {et~al.}(1990){Falomo}, {Melnick}, \& {Tanzi}}]{fal90}
{Falomo}, R., {Melnick}, J., \& {Tanzi}, E.~G. 1990, \nat, 345, 692

\bibitem[{{Falomo} {et~al.}(1994){Falomo}, {Scarpa}, \& {Bersanelli}}]{fal94}
{Falomo}, R., {Scarpa}, R., \& {Bersanelli}, M. 1994, \apjs, 93, 125

\bibitem[{{Fan} \& {Lin}(2000)}]{fan2000}
{Fan}, J.~H. \& {Lin}, R.~G. 2000, \apj, 537, 101

\bibitem[{{Ghisellini} {et~al.}(2010){Ghisellini}, {Tavecchio}, {Foschini},
  {Ghirlanda}, {Maraschi}, \& {Celotti}}]{ghisellini2010}
{Ghisellini}, G., {Tavecchio}, F., {Foschini}, L., {et~al.} 2010, \mnras, 402,
  497

\bibitem[{{Goldoni}(2011)}]{goldoni11}
{Goldoni}, P. 2011, Astronomische Nachrichten, 332, 227

\bibitem[{{Kacprzak} {et~al.}(2011){Kacprzak}, {Churchill}, {Barton}, \&
  {Cooke}}]{kac11}
{Kacprzak}, G.~G., {Churchill}, C.~W., {Barton}, E.~J., \& {Cooke}, J. 2011,
  \apj, 733, 105

\bibitem[{{Kotilainen} {et~al.}(1998){Kotilainen}, {Falomo}, \&
  {Scarpa}}]{kot98}
{Kotilainen}, J.~K., {Falomo}, R., \& {Scarpa}, R. 1998, \aap, 336, 479

\bibitem[{{Maraschi} \& {Tavecchio}(2003)}]{maraschi2003}
{Maraschi}, L. \& {Tavecchio}, F. 2003, \apj, 593, 667

\bibitem[{{Melnick}~J.(1994)}]{european1994emmi}
{Melnick}~J., {Dekker}~H., D.~S. 1994, EMMI \& SUSI: the ESO Multi-Mode
  Instrument and the Supurb Seeing Imager, ESO operating manual (ESO)

\bibitem[{{Pica} \& {Smith}(1983)}]{pica83}
{Pica}, A.~J. \& {Smith}, A.~G. 1983, \apj, 272, 11

\bibitem[{{Rector} \& {Stocke}(2001)}]{rector01}
{Rector}, T.~A. \& {Stocke}, J.~T. 2001, \aj, 122, 565

\bibitem[{{Sbarufatti} {et~al.}(2006){Sbarufatti}, {Treves}, {Falomo}, {Heidt},
  {Kotilainen}, \& {Scarpa}}]{sba06}
{Sbarufatti}, B., {Treves}, A., {Falomo}, R., {et~al.} 2006, \aj, 132, 1

\bibitem[{{Stickel} {et~al.}(1993){Stickel}, {Fried}, \& {Kuehr}}]{stickel93}
{Stickel}, M., {Fried}, J.~W., \& {Kuehr}, H. 1993, \aaps, 98, 393

\bibitem[{{Tavecchio} {et~al.}(2010){Tavecchio}, {Ghisellini}, {Ghirlanda},
  {Foschini}, \& {Maraschi}}]{tav10}
{Tavecchio}, F., {Ghisellini}, G., {Ghirlanda}, G., {Foschini}, L., \&
  {Maraschi}, L. 2010, \mnras, 401, 1570

\bibitem[{{Vernet} {et~al.}(2011){Vernet}, {Dekker}, {D'Odorico}, {Kaper},
  {Kjaergaard}, {Hammer}, {Randich}, {Zerbi}, {Groot}, {Hjorth}, {Guinouard},
  {Navarro}, {Adolfse}, {Albers}, {Amans}, {Andersen}, {Andersen}, {Binetruy},
  {Bristow}, {Castillo}, {Chemla}, {Christensen}, {Conconi}, {Conzelmann},
  {Dam}, {de Caprio}, {de Ugarte Postigo}, {Delabre}, {di Marcantonio},
  {Downing}, {Elswijk}, {Finger}, {Fischer}, {Flores}, {Fran{\c c}ois},
  {Goldoni}, {Guglielmi}, {Haigron}, {Hanenburg}, {Hendriks}, {Horrobin},
  {Horville}, {Jessen}, {Kerber}, {Kern}, {Kiekebusch}, {Kleszcz}, {Klougart},
  {Kragt}, {Larsen}, {Lizon}, {Lucuix}, {Mainieri}, {Manuputy}, {Martayan},
  {Mason}, {Mazzoleni}, {Michaelsen}, {Modigliani}, {Moehler}, {M{\o}ller},
  {Norup S{\o}rensen}, {N{\o}rregaard}, {P{\'e}roux}, {Patat}, {Pena}, {Pragt},
  {Reinero}, {Rigal}, {Riva}, {Roelfsema}, {Royer}, {Sacco}, {Santin},
  {Schoenmaker}, {Spano}, {Sweers}, {Ter Horst}, {Tintori}, {Tromp}, {van
  Dael}, {van der Vliet}, {Venema}, {Vidali}, {Vinther}, {Vola}, {Winters},
  {Wistisen}, {Wulterkens}, \& {Zacchei}}]{vernet11}
{Vernet}, J., {Dekker}, H., {D'Odorico}, S., {et~al.} 2011, \aap, 536, A105

\bibitem[{{Wills} {et~al.}(1992){Wills}, {Wills}, {Breger}, {Antonucci}, \&
  {Barvainis}}]{wills92}
{Wills}, B.~J., {Wills}, D., {Breger}, M., {Antonucci}, R.~R.~J., \&
  {Barvainis}, R. 1992, \apj, 398, 454

\end{thebibliography}
 \nocite{*}

\end{document}